\documentclass{iaus}
\usepackage{graphics}

  \checkfont{eurm10}
  \iffontfound
    \IfFileExists{upmath.sty}
      {\typeout{^^JFound AMS Euler Roman fonts on the system,
                   using the 'upmath' package.^^J}%
       \usepackage{upmath}}
      {\typeout{^^JFound AMS Euler Roman fonts on the system, but you
                   dont seem to have the}%
       \typeout{'upmath' package installed. iaus.cls can take advantage
                 of these fonts,^^Jif you use 'upmath' package.^^J}%
      }
  \else
  \fi

  \checkfont{msam10}
  \iffontfound
    \IfFileExists{amssymb.sty}
      {\typeout{^^JFound AMS Symbol fonts on the system, using the
                'amssymb' package.^^J}%
       \usepackage{amssymb}%

      }{}
  \fi

  \IfFileExists{amsbsy.sty}
    {\typeout{^^JFound the 'amsbsy' package on the system, using it.^^J}%
     \usepackage{amsbsy}}
    {\providecommand\boldsymbol[1]{\mbox{\boldmath $##1$}}}

\title[The ASTRA Spectrophotometer: A July 2004 Progress Report]{The ASTRA Spectrophotometer: A July 2004 Progress Report}

\author[Adelman, Gulliver, Smalley, Pazder, Younger, Boyd, \& Epand]%
{Saul J. Adelman$^1$%
  \thanks{ASTRA Contribution 2A. This work is supported by NSF grant AST-0115612 to The Citadel},
Austin F. Gulliver$^2$,
Barry Smalley$^3$,
John S. Pazder$^4$,
P. Frank Younger$^5$,
Louis J. Boyd$^6$,
\and Donald Epand$^6$}

\affiliation{{$^1$Department of Physics, The Citadel, 171 Moultrie Street, Charleston, SC 29409,
USA}\\
{$^2$Department of Physics \& Astronomy, Brandon University, Brandon, MB R7A 6A9, Canada}\\
{$^3$Astrophysics Group, School of Chemistry and Physics, Keele University, Straffordshire ST5 5BG, United Kingdom}\\
{$^4$Dominion Astrophysical Observatory, Herzberg Institute for Astrophysics, National Research of Canada, 5071 W. Saanich Road, Victoria, BC V9E 2E7, Canada}\\
{$^5$Aurora Astronomical Services, 585 Aurora Way, Victoria, BC V8Z 3J8, Canada}\\
{$^6$Fairborn Observatory, HCR5 2, Box 256, Patagonia, AZ, 85624,USA}}

\pubyear{2004}
\volume{224}
\pagerange{1--8}
\date{?? and in revised form ??}
\jname{The A-Star Puzzle}
\editors{J. Zverko, W.W. Weiss, J. Ziznovsky \& S.J. Adelman, eds.}

\begin{document}
\maketitle

\begin{abstract}
A cross-dispersed spectrophotometer with a back-illuminated uv-coated CCD detector and its automated 0.5-m telescope at the Fairborn Observatory, Washington Camp, AZ now under construction, should begin observations in 6 to 9 months. The Citadel ASTRA (Automated Spectrophotometric Telescope Research Associates) Telescope will be able to observe Vega the primary standard, make rapid measurements of the naked-eye stars, use 10 minutes  per hour to obtain photometric measurements of the nightly extinction, and obtain high quality observations of V= 10.5 mag. stars in an hour. The approximate wavelength range is $\lambda\lambda$3300-9000 with a resolution of 14 \AA~in first and 7 \AA~ in second order. Filter photometric magnitudes and indices will be calibrated in part for use as quality checks.

First a grid of secondary standards will be calibrated differentially with respect to Vega. They will also be used to find the nightly extinction. The candidates were selected from the most stable bright secondary stars of the grating scanner era, the least variable main sequence B0-F0 band stars in Hipparcos photometry, and metal-poor stars. Continued measurements of secondary stars will be used to improve the quality of the secondary standard fluxes. Science observations for major projects such as comparisons with model atmospheres codes and for exploratory investigations should also begin in the first year. The ASTRA team realizes to deal with this potential data flood that they will need help to make the best scientific uses of the data. Thus they are interested in discussing possible collaborations. In less than a year of normal observing, all isolated stars in the Bright Star Catalog that can be observed can have their fluxes well measured. Some A Star related applications are discussed.
\end{abstract}

\section{Introduction}

We are building a simple, efficient, and elegantly designed spectrophotometer with a CCD detector for use on an automated 0.5-m telescope at the Fairborn Observatory, Washington Camp, AZ.  This unique multiplexed instrument should produce high quality fluxes through at least the $\lambda\lambda$3300-9000 region for application to a wide variety of astrophysical problems.  With a resolution of 14 \AA ~in first and 7 \AA~in second order, one could synthesize indices for most optical region filter photometric systems and then perform studies using such indices.  But spectrophotometric data provides flux distributions in far greater detail, similar to classification spectroscopy.  Considerable astrophysical information that will be studied is lost at filter photometric resolutions. 

The telescope building has been completed and both the instrument and the telescope are well under construction.  The spectrophotometer will be thoroughly tested before being mated to the telescope. 
Their integration should require 3 to 6 months.  Regular observations are expected to begin in the first half of 2005.

Almost all the rotating grating scanners used for spectrophotometry and to measure the absolute fluxes of Vega and the secondary standards were retired well over a decade ago.  Most replacement instruments were not intended for stellar observations and lack the required accuracy and precision.  The stellar fluxes from rotating grating scanners may contain systematic wavelength-dependent errors due to those in the absolute calibration, extinction, bandpass centering, scattered light in the instrument, and other causes.  Such data typically consists of 15 to 20 values covering $\lambda\lambda$3400-7100 with 20 to 50 \AA~wide bandpasses usually at spectral regions with minimal line blanketing.  The extinction was based often on mean observatory values and the errors are rarely better than 1\%. Breger (1976a), Ardeberg \& Virdefors (1980), and Adelman et al. (1989), in particular, compiled these observations.  There are also 5-m Hale telescope measurements, based largely on rotating grating scanner calibrations, with now retired instruments: the multichannel spectrophotometer (Oke \& Gunn 1983, Gunn \& Stryker 1983) and the Double Spectrograph (Oke 1990 who provided standards for the Hubble Space Telescope) and satellite measurements of the optical ultraviolet (e.g., Code \& Meade 1979).  Many more stars had their ultraviolet fluxes measured to a reasonable accuracy and precision by the IUE satellite than by previous instruments in the optical region. 

With optical region grating scanner (and ultraviolet flux) data and Balmer line profiles, astronomers derived reasonably good effective temperatures and surface gravities of normal single slowly rotating B, A, and F stars.  These have been expressed in terms of filter photometric indices of various systems.  For stars with significant non-solar compositions, such calibrations are not necessarily accurate as metallicity, microturbulence, macroturbulence, and/or magnetic fields affect the stellar fluxes in subtle, yet measurable ways (Adelman \& Rayle 2000).  	

For stellar astrophysics, at the heart of our understanding the history and evolution of galaxies, this spectrophotometric instrument is critical for future advancements.  Spectrophotometry can also be an important technique for the study of solar system objects, nebulae, star clusters, and galaxies.  NEW DIRECTIONS IN SPECTROPHOTOMETRY (eds. Philip, Hayes \& Adelman 1988) discusses additional uses.  

We are not planning to improve the absolute flux calibration of Vega at this time.  Although our instrument would prove to be superior to the rotating grating scanners for this regard, a Fourier transform interferometer would be even better.

\section{The ASTRA Telescope}

The Citadel ASTRA telescope will be at Washington Camp, AZ, just north of the US-Mexico border about 30 kilometers east of Nogales, AZ.  The spectrophotometer will be mounted at the f/16 Cassegrain focus.  The science CCD is an Apogee Instruments Alta system with three-stage piezoelectric cooling.  The spectrophotometer will be in an insulated box, which will be kept at a temperature of 4$^{\circ}$ C.  The CCD is a E2V 30-11 with 1024 x 256 square 26-micron pixels and an ultraviolet coating.  Its temperature is expected to be kept close to -60$^{\circ}$C. By maximizing the optical ultraviolet response, the total exposure time to reach a minimal S/N ratio will be minimized for most stars.

Although the 0.5-m telescope will be the first of its type designed by Louis Boyd, it will incorporate many features used by the other small automated telescopes of Fairborn Observatory including control by ATIS (Automated Telescope Instruction Set whose documentation is available at www.fairobs.org), disk and roller drives on both axes, and a very short search and find interval of a few seconds on average between successive program stars.    Don Epand has already implemented most of the ATIS commands needed for spectrophotometric observations.  Flexure should not be a problem, but we will investigate this possibility during testing.
\section{The Spectrophotometer}

This instrument was designed to minimize flexure and the cost of construction.  The insulated case is a box, rectangular in cross-section.  The optical plate, made from 1.25 cm thick aluminum, will be attached to the telescope mounting collar with insulation to thermally isolate the telescope from the spectrophotometer.  The length overall will be roughly 38 cm,  greatest width 28 cm, and height 14 cm.  The case can be opened to provide access to the instrument.  Thermal control will be accomplished by using a 4$^{\circ}$C water supply and sufficient insulation.

The optical parts are secured in solid mounts.  With our slitless spectrophotometer, focusing with be done with the telescope.  The initial optimal optics placement on the optical plate uses a pinhole on the grating rotational axis.  The optics will be located empirically to provide instrumental alignment and focus. The grating rotation will be set at the initial alignment and thereafter only moved when absolutely necessary.  

A prismatic cross disperser provides sufficient order separation for the spectrograph to cover $\lambda\lambda$3000-10000 in a single exposure.  The main dispersion element is a 300 gr mm$^{-1}$ grating with a 8600 \AA~blaze.  From diffraction grating efficiency data, the optimal order coverage was found to be 5500 \AA~ to 10000  \AA~ in the first and 3000 \AA~ to 6000  \AA~ in the second order.  A 500 \AA~ overlap between the orders allows the crossover wavelength to be fine tuned based on the characteristics of the final grating.

A 1.0 arc second object is 2 pixels wide at the image, matching the image size to the Nyquist frequency of the 26 square micron CCD pixels.  The optical performance of the spectrograph at 80\% encircled energy is better than 17 microns (50\% in 8 microns) over the whole spectral range for a point source object.  A 1.0 arc second image of the star, which would have a width of 30 microns for a perfect camera, will have an image size, at worst, of 35 microns. The instrument has very low scattered light.  Baffles will be introduced if necessary.  As the smallest bandpasses can be two pixels wide, the resolution will be 14 \AA~ in the first and 7 \AA~ in the second order. 

To preserve the resolution set by the stellar image in this slitless design and to find cosmic ray hits, the spectrum will be widened to 5 or 6 pixels by a very slight mechanical rocking of the telescope.  The separation of the two orders is sufficient so that during the rocking, the sky measurements of each order will not overlap.  A 30 arcsec x 30 arcsec projected square hole in the 45$^{\circ}$ mirror used to acquire the star acts as a field stop.  As the CCD read noise is of order 8 electrons per pixel, rocking the spectrum does not significantly degrade the S/N ratio. 

The guide and centering camera optics are both standard off-the-shelf achromatic doublets.  For the guide camera the image scale is set to 2 pixels for a 1.0 arc second disc, and for the finding camera it is 3 pixels for a 1.0 arc second disc.  The mechanical and optical designs are being finalized  with measurements of the actual optical components.  

\section{Instrumental Diagrams}

We have four figures illustrating the optics of the ASTRA Spectrophotometer.  Figure 1 shows the positions of the dispersed light from the first and second orders on the CCD format.  Figures 2 and 3 follow the light paths as seen from the top and side of the instrumental enclosure.  Figure 4 is a spot diagram showing that the instrument produces images which fall well within a square whose side is twice one of the 26$\mu$ pixels.
\begin{figure*}
 \includegraphics{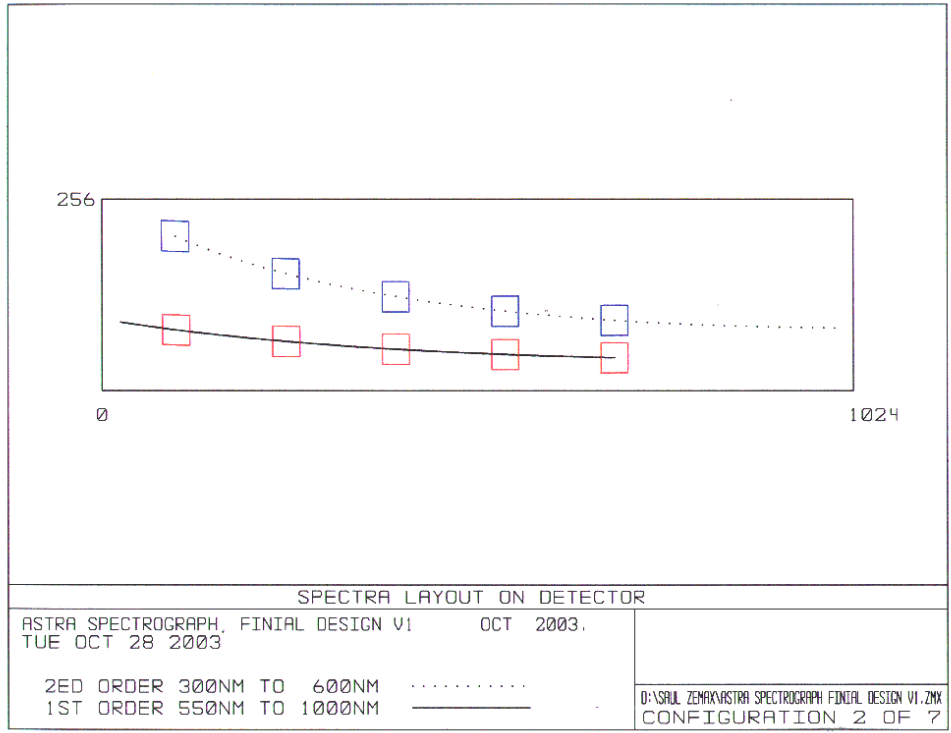}
  \caption{The CCD format showing the positions of the first and second orders.}
  \end{figure*}

\begin{figure*}
 \includegraphics{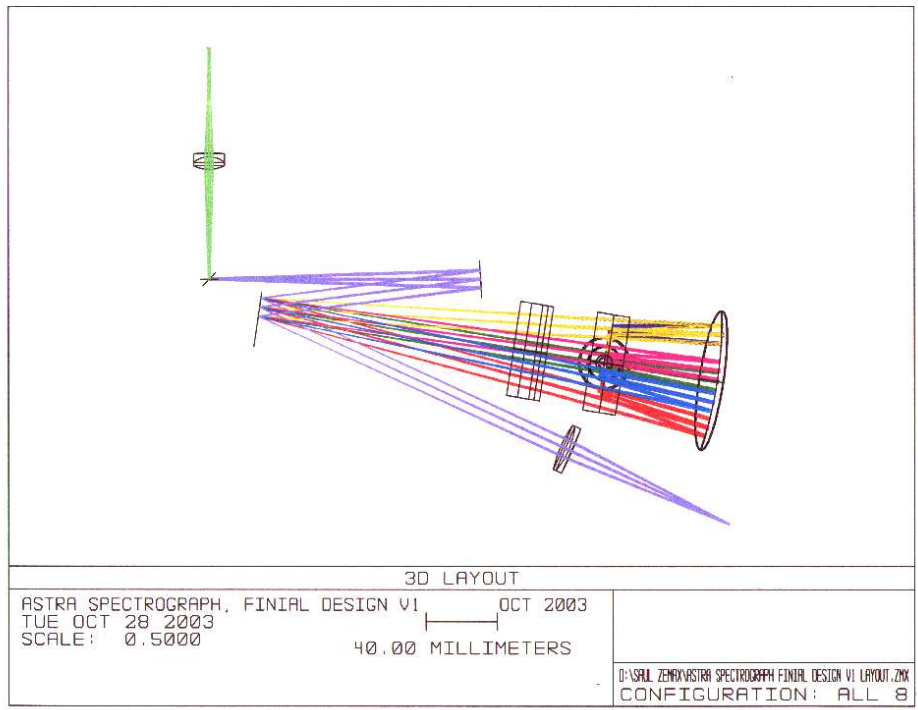}
  \caption{A view from the top of the spectrophotometer showing the light entering the instrument at the left, either reflecting off of the 45$^{\circ}$ field mirror, or passing through to the collimator, its dispersion by the grating and how the different orders are focused, and the use of the zeroth order light for guiding (lower right).}
  \end{figure*}

\begin{figure*}
 \includegraphics{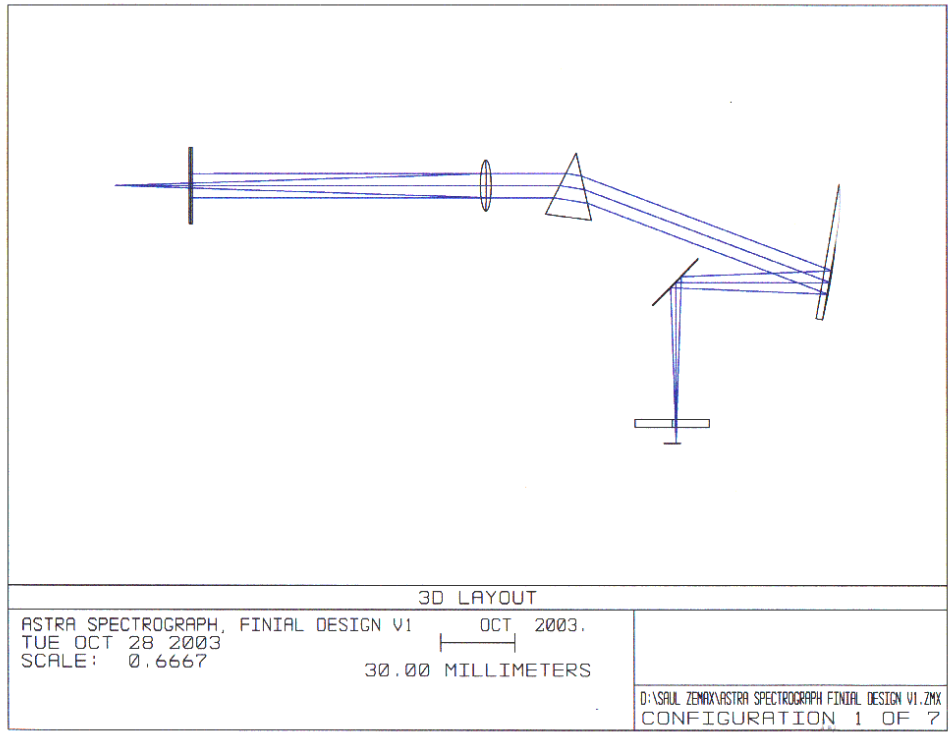}
  \caption{A side view of the light path. Here the light enters from the right, is reflected by the collimator and dispersed by the grating, then passes through the cross dispersing prism and imaged by the two mirror camera, which folds the light vertically to the science CCD.  The zeroth order and finder camera beam paths are not shown.}
   \end{figure*}

\begin{figure*}
 \includegraphics{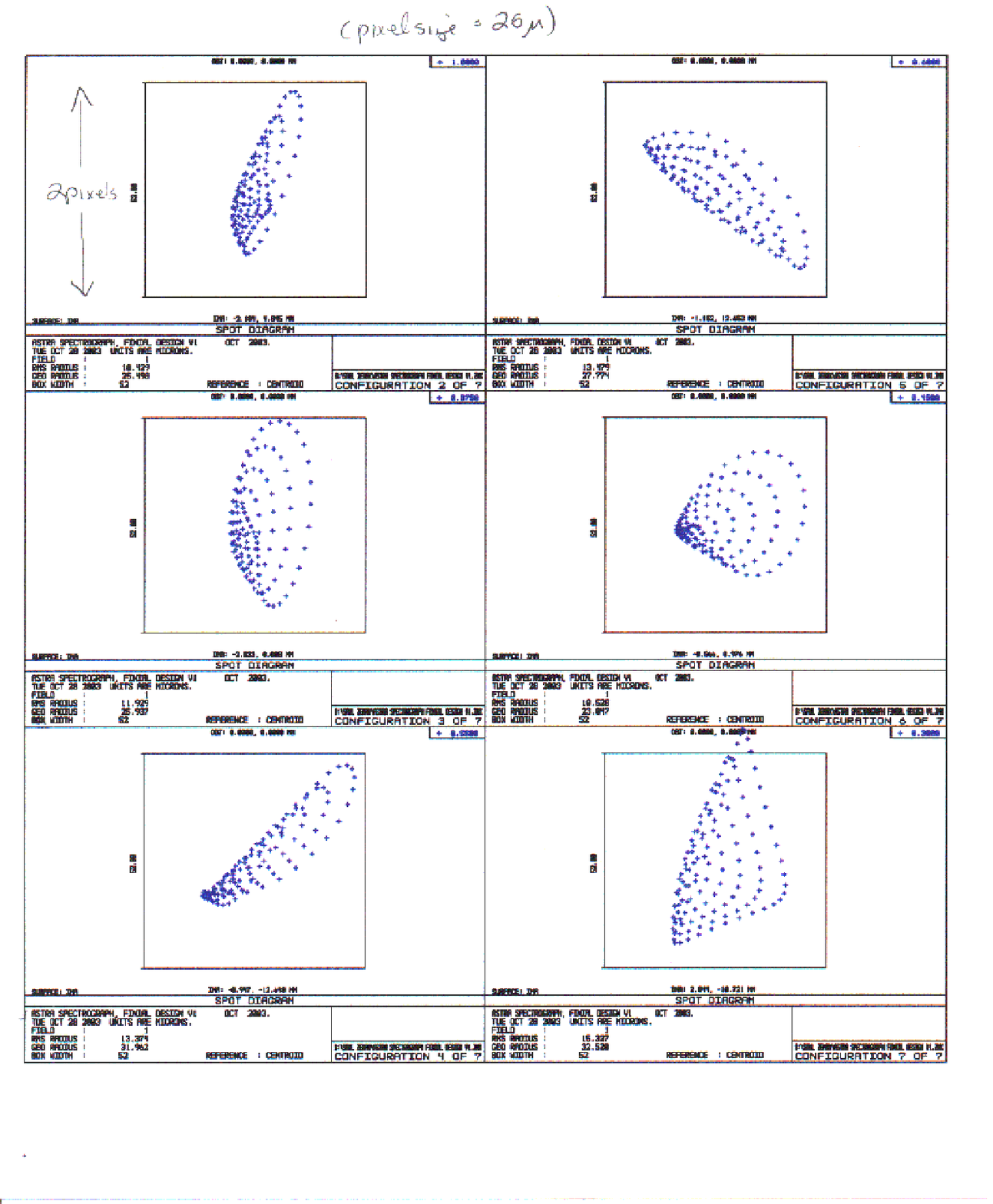}
  \caption{Spot diagrams for six wavelengths. These are the results of ray tracing calculations.  The square is two 26$\mu$ pixels on a side. }
  \end{figure*}

\section{Observing}

With our instrument's ability to multiplex measurements throughout its wavelength coverage, the time required to make frequent nightly observations of standards will be minimal.  To expose a 5th magnitude A0 V star and read the CCD should take about 25 seconds to obtain a signal-to-noise ratio of at least 200 at all useful wavelengths.  This estimate is based on the total system efficiencies, the T\"{u}g et al. (1977) calibration of Vega, a 0.5 m telescope, and allowance for atmospheric extinction.

The spectrophotometer uses two small CCD video cameras for acquiring stars and for guiding.  At the telescope focus, the flat mirror with the projected small hole is placed in the light beam at a 45$^{\circ}$ angle.  The light reflected from the mirror is focused on a small CCD camera for target acquisition.  Once the target star is found, the telescope moves automatically so that the light falls through the hole in the mirror, allowing the light to enter the spectrograph. The light from the zeroth order is focused onto a second small CCD camera.  The center of the image is kept at a particular pixel by use of an automated guider.  To obtain flat fields, the telescope will point to a screen on the role-off part of the building that is illuminated by quartz halogen lamps with non-uv excluding quartz envelopes sometime during the day in the dark shed.  The building doors will normally be locked when the lamps are on.
\section{Data Reduction}

We are writing our own computer codes to understand their details, include our experience with the site and the instrument, have the necessary flexibility to optimize them for our particular needs, and produce rigorous error estimates.  Gulliver and Graham Hill have written an optimal extraction computer program CCDSPEC to batch-reduce spectroscopic coud\'{e} CCD exposures that contain spectral data after being provided with the needed bias and flat exposures.  For the spectrophotometer a new version will reduce the first and second order spectra simultaneously.

Superior spectrophotometry requires measuring the instrumental profile and the scattered light that is found in pixels other than the one corresponding to its wavelength.  Although CCDs are ideally linear devices, their linearity and polarization properties must be checked.  This is important in observing Vega, which is much brighter than the other standards.  Our initial wavelength scale is the theoretical dispersion solution.  We will confirm it using strong features in the stellar fluxes such as lines of the Balmer and Paschen series and the Ca II K line.	

We are concerned about properly calibrating spectral regions containing the Balmer and Paschen lines and jumps and with large amounts of telluric features.  Smalley's numerical simulations show that one can recover the flux to better than 1\% in all but the regions of most severe telluric contamination.  Extinction modeling (see e.g. Hayes \& Latham 1975) will improve as we gain knowledge of the variations with time, airmass, and wavelength at the observing site. 

The raw data will be first run through a program to perform a constant extinction correction using all the data from the night.  If necessary the correction will be rerun with an appropriate time-interpolation scheme.  Once this is satisfactorily accomplished, the data will be transformed to absolute fluxes and error estimates will be made.

We want to obtain closure on the secondary standards by the end of the first year of observations and then concentrate on the other projects to demonstrate the quality of the data and the various astrophysical uses of spectrophotometric data.  We will make extensive comparisons with existing spectrophotometry.

\section{Projects}

During the first two years of observing, we will begin two major projects:

{\it 1. Revision and Extension of the Secondary Standards:} The fluxes must be reduced to a uniform system based on the absolute fluxes of Vega.  Not all possible secondary standards are equally good for calibrating all wavelength regions.  Taylor (1984) extended and made more uniform the existing bright star standards.  For our instrument, they must be redetermined as its resolution is greater than what had been the usual practice, of order 25 \AA.  Standards for larger telescopes brighter than our faint limit will also be checked and extended.  We selected some 250 candidates for secondary stars mainly from the least variable B0 to F0 stars in Hipparcos photometry to supplement those given by Taylor.  The measurements of the secondary stars for extinction will also be used to improve the quality of our secondary standard fluxes.

{\it 2. Sample Fluxes of Population I and II Stars:} This longer-term project will enable population synthesis analyses which require high quality optical region fluxes of all known types of stars.  We will observe all single stars in the Bright Star Catalog and its supplement, stars in clusters and associations that pass within 45$^{\circ}$ of our zenith and nearby stars with good distances to empirically define the Zero Age Main Sequence and calibrate the HR diagram.  The latitude of Fairborn Observatory is approximately  N31.5$^{\circ}$. Two important auxiliary projects are: 

{\it A. Comparisons with Model Atmospheres:} Model atmospheres analytically link the physical properties of stars (M, R, L, and composition) and the observed flux distribution and spectral line profiles.  By comparing predictions of model atmospheres with spectrophotometric fluxes (and Balmer line profiles) effective temperatures and surface gravities can be found for a wide variety of stars. Our data should be far superior to existing data for this purpose.  Comparisons between the best-fitting model atmospheres calculated with different codes for the same star can be performed to give insight into how well each code reproduces these observables.  By also deriving the elemental abundances, consistency checks can be made.  Hill, Gulliver \& Adelman (1996) developed a powerful fitting program STELLAR which now uses a four-dimensional grid of SYNTHE synthetic spectra (Kurucz \& Avrett 1981) and ATLAS9 model atmospheres (Kurucz 1993) continuous energy distributions to perform a simultaneous rms error fit to the observed metallic and hydrogen line spectra as well as the continuous flux distribution of stars. We have calculated extensive grids of ATLAS9 model atmospheres for B, A, and F stars with a variety of chemical compositions and microturbulences.  As a complementary project we plan to obtain Balmer profiles of  spectrophotometric targets at the Dominion Astrophysical Observatory's 1.22-m telescope.   

For parts of the HR diagram for which codes other than ATLAS9 produce the best model atmospheres, we would like to work with their authors and users.  We recognize the need to include effects of sphericity, NLTE, and velocity fields for some types of stars.

{\it B. Synthetic Colors and Line Indices from Spectrophotometry: } By synthesizing colors from the spectrophotometry one can check for consistency with photometry and/or provide photometric indices for stars that lack such values (Breger 1976b).  Systems of particular interest include Johnson UBV, Cousins RI, Sloan, $\Delta$a, Str\"{o}mgren, Geneva, and Vilnius/Str\"{o}mvil. That our instrument will produce data without major gaps in wavelength coverage will be a significant advantage over most previous scanner flux data for color synthesis. 

The strongest metal lines can be seen in continuous wavelength spectrophotometry at somewhat lower resolution (e. g. 20 \AA, Fay et al. 1973).  Hence one will be able to create many line strength measures such as use the Ca II K line to assess the metallicity of many stars.  Further for cool stars, one could measure the dependence of strong spectral features as functions of surface, gravity, and [Fe/H] as have Burstein et al. (1986) and Gorgas et al. (1993) and use them to study Population II objects.
\section{Opportunities for Collaboration}

The Fairborn Observatory site usually has the equivalent of about 140 photometric nights per observing season from the middle of September to the beginning of July.  With a telescope and instrument whose declination range for useful operation is N80$^{\circ}$ to  S35$^{\circ}$, we hope to obtain several 10s of thousands of observations per year.  Even using automated fitting routines, it is beyond the combined abilities of Adelman, Gulliver, and Smalley, who will be planning the observing and reducing the data to a usable form, to successfully analyze more than a small fraction of the potential observations.  As they realize that they will need help to make the best scientific uses of the ASTRA data, they are interested in finding collaborators.  

The two major projects discussed above are basic to other applications and are natural parts/products of the calibration effort.  The auxiliary projects can be done in collaboration with others.  The comparison of observations with model atmospheres requires the calculation of grids of model atmospheres as well as observations of Balmer line profiles.  Observations of H$\alpha$, H$\beta$, H$\gamma$, and perhaps other hydrogen lines at high dispersion will help us understand the usefulness of those at spectrophotometric resolution and vice versa.  The calculation of synthetic colors and line indices for all useful systems will become part of our data reduction effort.

We are particularly interested in using the ASTRA system to find stellar parameters and to investigate physical processes in stellar atmospheres.   Spectrophotometry will be especially useful to study those kinds of stars whose energy distributions change shape.  For variable stars it is important to realize that the time between successive observations can be as short as of order 30 seconds for stars with V magnitudes in the 3.5 to 5 mag. range.   Experimental programs to investigate all kinds of stars are appropriate.

As many studies of variable stars will likely utilize local spectrophotometric standards, we want to calibrate such stars as part of the initial effort.  Thus we are interested in making some arrangements for collaboration before the start of observations.  We are also very interested in analysis tools that will be useful for many different projects.  As we are using LINUX and UNIX platforms to run the various applications for ASTRA, we would like such tools to run on these platforms.

Many possible A Star projects involve determinations of T$_{\rm eff}$ and log g or searches for companions. Some examples of other projects follow

A-type supergiants : sphericity effects, and pulsational properties

Am stars  :  determining where convection becomes important and which theory is most appropriate, searches for broad, continuum features

HgMn stars : searches for signatures of peculiar compositions, the study of broad, continuum
 	features

Magnetic CP stars : variability around rotational periods, searches for secondary
 	periods, details of broad, continuum features, use with Doppler imaging studies,
	long term changes in variability characteristics

RR Lyrae stars : pulsational properties and searches for hydrodynamical signatures

\section{Information for Possible Collaborators}

1. All collaboration teams will have Adelman, Gulliver, and/or Smalley as a member(s) and as a coauthor(s) on all resulting papers.  Two or all of them may participate in a given project if they are particularly interested and contribute more than data and an understanding of its properties. 

2. Page charges, if any, will usually be the responsibility of the other collaborators. 

3. All papers will be part of an ASTRA paper series.

4. The spectrophotometric values will appear in a public catalog after a substantial usage of the data has been made.  This catalog, coauthored by Adelman, Gulliver, and Smalley, will include references to papers in which the data were used.

5. The data will be analyzed and submitted for publication in a timely manner.  To keep ASTRA operations funded requires the publication of scientific important and useful results.

6. We are willing to obtain simultaneous as well as phase-dependent observations.

7. Observations may be used for more than one program.

8. We will endeavor to provide sufficient observations for each accepted collaboration in the first 1.5 years of scientific operations so that an assessment of the project can be performed and an initial publication can be written and submitted.

9. We are interested in long term partnerships.

If you are interested, please contract one of us as soon as possible.  Our email addresses follow:

Saul J. Adelman   	(adelmans@citadel.edu),

Austin F. Gulliver  	(gulliver@brandonu.ca), and 

Barry Smalley   	(bs@astro.keele.ac.uk)

\begin{acknowledgments}

The ASTRA spectrophotometer is supported by grant AST-0115612 from the United States National Science Foundation to The Citadel.  Additional funding has been provided by The Citadel Foundation.
\end{acknowledgments}

\end{document}